\documentclass[elsart12,epsf,eqsecnum,graphics,cite]{revtex4}
\def\beq{\begin{equation}}

\def\eeq{\end{equation}}

\usepackage{epsfig,epsf}
\usepackage{graphicx}
\usepackage{grffile} 

\usepackage{amsmath,amsfonts,amssymb,amsthm,nccmath,latexsym,mathtools} \usepackage[mathscr]{euscript}

\newcommand{\be}{\begin{equation}}
\newcommand{\ee}{\end{equation}}

\begin{document}

 
\voffset1.5cm

\title{Photons without vector fields}
\author{Ibrahim Burak Ilhan$^{1}$ and Alex Kovner$^{1}$}

\affiliation{
$^1$ Physics Department, University of Connecticut, 2152 Hillside
Road, Storrs, CT 06269-3046, USA}

\begin{abstract}
In this note we continue to pursue the question whether gauge theories can be represented in terms of effective ``scalar'' degrees of freedom. We provide such a consistent representation for a free photon theory in 3+1 dimensions. Building on results of \cite{biz} we construct a Lagrangian with a four derivative kinetic term, and demonstrate that despite seeming nonlinearity of the theory it is equivalent to a theory of a free photon. 
\end{abstract}
\maketitle
\section{Introduction}

In a recent paper \cite{biz} we constructed an effective theory of  scalar fields in 3+1 dimensions with certain features that could potentially mimic the low energy limit of gluodynamics. The idea was based on the known relation between confinement and $Z_N$ symmetry in 2+1 dimensions \cite{zn,abelnonabel}. 
Although the main aim of this construction was to explore possible low energy representation of nonabelian theories, a certain limit of the construction should encompass an abelian gauge theory, and its simplest limit - the theory of a free photon \cite{abelnonabel}.  

The construction of \cite{biz} despite having some useful features failed to describe exactly this abelian limit. 
In this note we rectify this problem. We modify the abelian limit of model of \cite{biz}, and demonstrate that the modified model is exactly equivalent to  a theory of the free photon.


The abelian limit of the model of \cite {biz} can be written as

\be \mathcal{L} =\frac{1}{4} f_{\mu\nu}f^{\mu\nu}= -\frac{1}{2}(\vec{e}^2 - \vec{b}^2)\ee

with $e_i = f_{i0}$ and $b_i = \frac{1}{2}\epsilon_{ijk}f_{jk}$ ($f_{ij}=\epsilon_{ijk}b_k$); and $f^{\mu\nu}$ defined as:

\be f^{\mu\nu}=g\epsilon^{\mu\nu\alpha\beta}\epsilon^{abc}\phi_a\partial_\alpha \phi_b \partial_\beta \phi_c \ee
where  the scalar fields $\phi_a; \ \ a=1,2,3$ are constrained by $\phi^a\phi^a=1$.
This model was previously discussed in \cite{eduardo}. There as well as in \cite{biz} it was shown that despite some similarities,  it fails to describe a free photon primarily because the magnetic field is not required to be divergenceless $\partial_ib_i\ne 0$. 

To rectify this problem, we now consider the following setup

\be \mathcal{L} =\frac{1}{4} F_{\mu\nu}F^{\mu\nu}= -\frac{1}{2}(\vec{E}^2 - \vec{B}^2)\label{lagrangian}\ee

\be F^{\mu\nu}=g\epsilon^{\mu\nu\alpha\beta}[\epsilon^{abc}\phi_a\partial_\alpha \phi_b \partial_\beta \phi_c + (n \cdot \partial) n_\alpha \partial_\beta \Phi]\ee
where $n= \left(1,0,0,0\right)$ is a timelike unit vector and  $\Phi$ is an additional scalar field \cite{gliozzi}.

The presence of an explicit timelike vector seems to render the model non Lorentz invariant. However, as we will see below this is not quite the case. The model does 
possess a Lorentz invariant superselection sector, and it is this sector that is equivalent to QED.

In this note we discuss how this modification changes the model in the Abelian limit. We will show that  the theory has the same canonical structure as free electrodynamics. This includes the commutation relations between ``electric" and ``magnetic" fields as well as the Hamiltonian. The model is therefore equivalent to a theory of a free photon, even though it is not formulated in terms of the vector potential.
We also discuss the action of Lorentz transformations on the basic degrees of freedom of the model. We show explicitly that the fields $\phi^i$ are not covariant scalar fields, but rather have an ``anomalous" term in their Lorentz transformation law. With this modification we show that the model is indeed Lorentz invariant.

\section{Equations of Motion and Canonical Structure}
\subsection{Equations of motion.}

Let us define two independent fields $\chi$ and $z$ via  $\phi_3 = z$ and $\phi_1 + i \phi_2 = \sqrt{1 - z^2}e^{i\chi}$. The  electromagnetic field can now be written as:
 
 \be F^{\mu\nu} =  \epsilon^{\mu\nu\alpha\beta}[-2 \partial_\beta \chi \partial_\alpha z + n_\alpha \partial_\beta \partial_0 \Phi ] \label{cons}\ee
or component by component 
\be E_i  = 2   \epsilon_{ijk}\partial_j z \partial_k \chi \label{elek}\ee
\be B_k = [2 (\partial_k \chi \partial_0 z - \partial_0 \chi \partial_k z ) - \partial_k \partial_0 \Phi] \label{mag}\ee
 



The Lagrangian equations of motion that follow from the Lagrangian eq.(\ref{lagrangian}) are

\be \partial_0 \partial_k \left[F_{ij}\epsilon^{ij0k}\right]=0 = \partial_0 \partial_k B_k \label{l1}\ee


\be \partial_\beta \chi \partial_\alpha (F_{\mu\nu}\epsilon^{\mu\nu\alpha\beta})=0 = \partial_k \chi \partial_\alpha (F_{\mu\nu}\epsilon^{\mu\nu\alpha k}) = \partial_k \chi (\partial_0 B_k + (\partial \times E)_k) \label{l2}\ee
\be \partial_\beta z \partial_\alpha (F_{\mu\nu}\epsilon^{\mu\nu\alpha\beta})=0 = \partial_k z \partial_\alpha (F_{\mu\nu}\epsilon^{\mu\nu\alpha k }) = \partial_k z (\partial_0 B_k + (\partial \times E)_k)\label{l3}\ee

Eq.(\ref{l1}) is a local conservation equation of a ``magnetic charge density" $\partial_kB_k$. It ensures that the Hilbert space of the theory is divided into ``superselection sectors" with fixed value of the magnetic charge density. In order to preserve translational invariance we limit ourselves to the sector with $\partial_k B_k=0$. Our considerations in the rest of this paper pertain to this superselection sector alone.

Using this constraint on the magnetic field,  equations eq.(\ref{l2})  and eq.(\ref{l3}) can be inverted, with the result\footnote{ Strictly speaking there is an ambiguity in the inversion of eqs.(\ref{l2},\ref{l3}). The general solution is 
$\partial_0 B_k + (\partial \times E)_k = \alpha E_k$ with an arbitrary constant $\alpha$. 
However, given that   $B$ and $\partial \times E$ are pseudo vectors while  $E$ is a vector, a non vanishing value of $\alpha$ would violate parity. Requiring parity invariance of the equations resolves the ambiguity and sets $\alpha=0$.}
\be \partial_0 B_k + (\partial \times E)_k = 0\ee

Recall that with the field strength components given by eq.(\ref{cons}),  the equation 
\be \partial_\mu F^{\mu\nu}=0\ee
is satisfied identically. We thus  have the full set of Maxwell's equations.

\subsection{The Hamiltonian.}
We now demonstrate that the Hamiltonian and the canonical commutation relations of the electromagnetic fields in our model are identical to those in pure QED.

The canonical momenta can be calculated from equation (\ref{cons}) as :

\be p_z = \frac{\delta L}{\delta \partial_0 z} =   F_{ij}\epsilon^{ij0k}\partial_k \chi = 2   B_k \partial_k \chi= 2   \partial_k \chi [2  (\partial_k \chi \partial_0 z - \partial_0 \chi \partial_k z) - \partial_k \partial_0 \Phi]\label{pz}\ee

\be p_\chi = \frac{\delta L}{\delta \partial_0 \chi} =    F_{ij}\epsilon^{ijk0}\partial_k z = -2   B_k \partial_k z= -2   \partial_k z [2  (\partial_k \chi \partial_0 z - \partial_0 \chi \partial_k z) - \partial_k \partial_0 \Phi]\label{pchi}\ee

\be p_\Phi = \frac{\delta L}{\delta \partial_0 \Phi} = \frac{1}{2} \partial_k (F_{ij}\epsilon^{ij0k}) =  \partial_k B_k =  \partial_k [2 (\partial_k \chi \partial_0 z - \partial_0 \chi \partial_k z) - \partial_k \partial_0 \Phi] \label{fidot}\ee
It is a straightforward matter to express the time derivative of $\chi$ and $z$ as: 

\be  \dot{\chi} = \frac{1}{E^2}[p_z (z \chi) + p_\chi \chi^2 +  \epsilon_{ijk}\dot{\Phi}_i E_j \chi_k] \label{chidot}\ee
\be \dot{z} = \frac{1}{E^2} [p_z z^2+ p_\chi (z \chi) +  \epsilon_{ijk}\dot{\Phi}_i E_j z_k] \label{zdot}\ee
The time derivative of $\Phi$ is related to canonical momenta via: 

\be p_\Phi 
= \partial_k \left[ \frac{1}{ E^2} \epsilon_{klm}E_l \left(p_z z_m + p_\chi \chi_m\right)- \frac{1}{E^2} E_k E_i \dot{\Phi}_i\right] 
\ee
or in terms of a ``vector potential"
\be A_k = \frac{1}{E^2} \epsilon_{klm}E_l \left( p_z z_m + p_\chi \chi_m \right)\ee
as
\be p_\Phi = \partial_k \left(  A_k - \hat{E}_k\hat{E}_i \dot{\Phi}_i\right)\label{pphi}\ee

The Hamiltonian is then calculated as:

\be H = \int d^3x \left[p_z \dot{z} +p_\chi \dot{\chi} + p_\Phi \dot{\Phi} - L \right]= \int d^3x \frac{1}{2} \left( E^2 + B^2\right) \label{hamil}\ee

where we have neglected a  boundary term $ \int d^3x\partial_k \left(B_k \dot{\Phi}\right)$. 

\subsection{Canonical structure}
To show that our model is equivalent to QED we need to make sure that the canonical commutation relations of $E_i$ and $B_i$ are identical in the two theories.

First off all, since all components of the electric field in our model are functions only of coordinates and not canonical momenta, they commute with each other

\be \left[ E_i(x),E_j(y)\right]=0\ee

Our next goal is to calculate the commutator between electric and magnetic field. In order to do that, we set
 $p_{\Phi}=0$, as we are only interested in this super selection sector of the theory. Then, eq.(\ref{pphi}) becomes

\be  \frac{\partial_k A_k}{ E}  =  \hat E_k \partial_k \left(\frac{\hat{E}_i \dot{\Phi}_i}{E}\right)\ee
where we have used $\partial_kE_k=0$.

The formal solution of this equation can be obtained as 
\be \hat{E}_i \dot{\Phi}_i= E(x)\int^x_{-\infty}dl_C \frac{\partial_k A_k}{ E}\ee
where the integral is along the contour $C$ which  starts at $x$ and goes to infinity (boundary of space). The contour is everywhere parallel to the direction of the electric field.

 Using the definition, we have:

\be B_k =  A_k -  E_k \int^x_{-\infty} dl_C \frac{\partial_m A_m}{E}\label{bk}\ee

As an intermediate step for the calculation of the commutator  $[E,B]$ we consider

\begin{align}\begin{split}\left[ E_i (x), A_{k}(y)\right] &=2i \frac{E_l(y)}{E^2(y)} \epsilon_{iab}\epsilon_{klm}\left[\partial_a^x\delta(x-y) \chi_b(x)z_m(y) + \partial_b^x \delta(x-y)z_a(x)\chi_m(y)\right] \\
&=2 i  \frac{E_l(y)}{E^2(y)}\epsilon_{iab}\epsilon_{klm}\partial_a^x \delta(x-y)[\chi_b(y)z_m(y) - z_b(y)\chi_m(y)]\\
&=i\hat{E}_l(y)\hat{E}_c(y) \epsilon_{iab}\epsilon_{klm}\epsilon_{cmb}\partial_a^x \delta(x-y)
=i\left[ \epsilon_{iak} - \hat{E}_b(y)\hat{E}_k(y)\epsilon_{iab}\right]\partial_a^x\delta(x-y)\\
\label{ea} \end{split}\end{align}

Using this, we can calculate
 
\begin{align}\begin{split}\label{split}
\left[ E_i (x), B_{k}(y)\right] 
&=[E_i(x), A_k(y)] - E_{k}(y)\int^y_{-\infty}dl_C \frac{\partial^t_m \left[E_i(x), A_{m}(t)\right]}{E(t)} \\
&=[E_i(x), A_k(y)] - E_{k}(y)  \int^y_{-\infty} dl_C\frac{1}{E(t)}\partial_m^t \left[ \left(\epsilon_{iam} - \hat{E}_b(t)\hat{E}_m(t) \epsilon_{iab}\right)\partial_a^x\delta(x-t)\right]\\
&=[E_i(x), A_k(y)]  +E_{k}(y) \int^y_\infty dl_C \hat{E}_m(t)\partial_m^t\left(\frac{\hat{E}_b(t)}{E(t)} \epsilon_{iab} \partial_a^x \delta(x-t)\right)\\
&=i\epsilon_{iak}\partial_a^x \delta(x-y)\\
\end{split}\end{align}

where we have used the fact that the integration contour $C$ is  defined to run in the direction of  electric field, and have assumed that the fields decrease fast enough at the boundary. 

The commutator eq.(\ref{split}) coincides with the corresponding commutator in QED.

We now turn to the commutator of components of magnetic fields. 

It is straightforward to show that  $[B_i(x), B_a(y)]=0$  as long as the curve $C_x$ that defines $B_i(x)$ in eq.(\ref{bk}) does not contain the point $y$, and $C_y$ does not contain $x$.
When this condition is not met, the direct calculation of the commutator is not straightforward. Instead of attempting it, we take an indirect way. A set of relations that involve the commutator in question are easily obtained. Consider for instance
\be [B_i(x)\partial_i \chi(x), B_j(y)\partial_j z(y)] = [p_z(x), p_\chi (y)]=0  \ee
Trivially:
\begin{align} &B_i(x) \partial_j z(y)[\partial_i \chi(x), B_j(y)]+ B_j(y)\partial_i\chi(x)[B_i(x),\partial_j z(y)]+ \partial_i \chi(x)\partial_j z (y)[B_i(x), B_j(y)]=\nonumber
\\&  \left( B_i(x)\partial_j z(y) \partial_i^{(x)}\frac{\partial A_j(y)}{\partial p_\chi(x)} -  B_j(y)\partial_i \chi(x) \partial_j^{(y)}\frac{\partial A_i(x)}{\partial p_z(y)}\right) + \partial_i \chi(x)\partial_j z (y)[B_i(x), B_j(y)]=\nonumber
\\&(B_i(y)\partial_i^{(x)}\delta(x-y) + B_i(x) \partial_i^{(y)}\delta(x-y)) + \partial_i \chi(x)\partial_j z (y)[B_i(x), B_j(y)]=\nonumber
\\& \partial_i \chi(x)\partial_j z (y)[B_i(x), B_j(y)]=0\label{bebe1}
\end{align}

Here we used the fact that $E_kz_k=E_k\chi_k=0$  and the constraint $\partial_i B_i=0$. Similarly 

\be \partial_i z (x)\partial_j z (y)[B_i(x), B_j(y)] = \partial_i \chi(x)\partial_j \chi (y)[B_i(x), B_j(y)]=0 \label{bebe2}\ee

And by $\partial_k B_k = 0$, we have:
\be
\partial_i z (x)\partial^{y}_j  [B_i(x), B_j(y)] = \partial_i \chi (x)\partial^{y}_j  [B_i(x), B_j(y)] = \partial^{x}_i \partial^{y}_j  [B_i(x), B_j(y)] =0 \label{bebe3}
\ee
Thus the commutator matrix $M_{ij}(x,y)\equiv  [B_i(x), B_j(y)] $, antisymmetric under the exchange  $(i,x) \leftrightarrow (j,y)$ satisfies the set of equations  eqs.( \ref{bebe1}-\ref{bebe3}). 
The general solution for these equations is given by
\be M_{ij}(x,y) = E_i(x)F_j(y) - E_j(y)F_i(x)\ee
where $F_i(x)$ is an arbitrary function.
However, we have already established that if $x$ does not belong to $C_y$ and $y$ does not belong to $C_x$, then $M_{ij}(x,y)=0$. This unambiguously fixes $F_i(x)=0$, so that we have:
\be [B_i(x), B_j(y)]=0\ee
for all $x,y$.

\section{Lorentz Transformations of the Fields} 
The final point we address is the Lorentz transformation properties of the fields $z$ and $\chi$. Since the electric and magnetic field are covariant components of the 
Lorentz tensor, it is clear that $z$ and $\chi$ cannot be covariant scalar fields. The transformations of $z$ and $\chi$ under rotations are the same as those of covariant fields, and we will not deal with those here. 

Let us parametrize the infinitesimal Lorentz transformation properties of these fields in the following way:

\begin{align}\begin{split}z(x) &\rightarrow z(\Lambda^{-1} x) = (1+\beta \Delta)z(x) + a\\
\chi(x) &\rightarrow \chi(\Lambda^{-1} x) = (1+\beta \Delta)\chi(x) + b\\
\Theta(x) \equiv \partial_0\Phi&\rightarrow \Theta (\Lambda^{-1} x) = \Theta(x) + c\end{split}\label{LT}\end{align}
Here $\beta$ is the boost parameter and $ \Delta \equiv {\omega^\mu}_\nu x^\nu \partial_\mu$ with $\omega^\mu_\nu$  - an antisymmetric generator of Lorentz transformation. In particular  for a boost in the direction of a unit vector $\hat n$, $\omega^i_0= \hat{n}_i$. The noncanonical terms 
 $a,b$ and $c$ are to be determined such that $F^{\mu\nu}$ transforms as a tensor.
 
For simplicity, let us consider explicitly a boost transformation in the first direction, $\hat n=(1,0,0)$. The transformation of the components of the field strength tensor are
\be
E_2(x) \rightarrow E_2(\Lambda^{-1} x) - \beta B_3 (\Lambda^{-1} x) 
\ee

on the other hand, writing this in terms of $z$, $\chi$ and $\Theta$  we have:
\be E_2(x) = 2  [ \partial_3 z(x) \partial_1 \chi(x) - \partial_1 z (x) \partial_3(x)] \rightarrow 2  [ \partial_3 z(\Lambda^{-1} x) \partial_1 \chi(\Lambda^{-1} x) - \partial_1 z (\Lambda^{-1} x) \partial_3(\Lambda^{-1} x)]\ee


Equating the two we obtain:
\be  - \beta \partial_3 \Theta + 2[ \partial_3 z \partial_1 b +\partial_3 a \partial_1 \chi - \partial_1 z \partial_3 b- \partial_1 a \partial_3 \chi] =0\ee

Similarly  by considering the transformation of   $E_1$ we obtain

\be 2 ( \partial_2 z \partial_3 b + \partial_2 a \partial_3 \chi - \partial_3 z \partial_2 b - \partial_3 a \partial_2 \chi)=0
\ee

and for $E_3$:

\be \beta \partial_2 \Theta + 2 [\partial_1 z \partial_2 b + \partial_1 a \partial_2 \chi - \partial_2 z \partial_1 b - \partial_2 a \partial_1 \chi] =0\ee

Defining for convenience  $f_i = 2  (a \partial_i \chi - b \partial_i z )$, and $u_i = (0, \beta \partial_3 \Theta, -\beta \partial_2 \Theta)$ the above equations can be written as
\be \epsilon_{ijk}\partial_j f_k = u_i\ee
The general solution for   $f$ is:
\begin{align}\begin{split} f_i &= -\frac{\epsilon_{ijk}\partial_j u_k}{\partial^2} +\partial_i \tilde\lambda \\
&=\beta\hat n_i \Theta + \partial_i \lambda \label{fv}
\end{split}\end{align}
where 
\be \tilde\lambda - \beta\frac{\hat n_i \partial_i}{\partial^2}\Theta = \lambda  \ee
where the function $\lambda$ still has to be determined.

We can now solve eq.(\ref{fv}) for $a$ and $b$ by noting that eq.(\ref{fv}) is identical to eq.(\ref{mag})  with the substitution
\begin{align}\begin{split}
&  \partial_0 z \rightarrow a\\
& \partial_0 \chi \rightarrow b\\
&\partial_0 \Phi \rightarrow \lambda\\
& B_k \rightarrow \beta\Theta \hat n_k
\end{split}\end{align}

Using eqs (\ref{chidot}, \ref{zdot}) we find:

\begin{align}\begin{split}
a&=  \frac{1}{E^2}(\beta\Theta \hat n_i + \lambda_i)\epsilon_{ijk}E_j z_k\\
b&= \frac{1}{E^2}(\beta\Theta \hat n_i + \lambda_i)\epsilon_{ijk}E_j\chi_k
\label{ab}\end{split}\end{align}

With this, eq. (\ref{fv}) yields the equation for $\lambda$:
\be E_i(\beta\Theta\hat n_i+ \partial_i \lambda )=0 \ee
from which we get:
\be \lambda(x) = -\beta \int^x_\infty dl_C \hat{E}_i \hat n_i\Theta \label{lambda}\ee
where again, $C$ is a curve in the direction of $E$

To determine the remaining function $c$ we consider the transformation of the magnetic field
For the transformation of the magnetic field, we have:

\begin{align}\begin{split}
B_1(x) &\rightarrow B_1(\Lambda^{-1}x) = 2 [\partial_1 \chi(\Lambda^{-1}x)\partial_0 (\Lambda^{-1}x) - \partial_0 \chi(\Lambda^{-1}x) \partial_1 z (\Lambda^{-1}x)] - \partial_1\Theta(\Lambda^{-1}x)\end{split}\end{align}
which yields
\be
2 [ \partial_1 \chi \partial_0 a + \partial_1 b \partial_0 z - \partial_0 \chi \partial_1 a -\partial_0 b \partial_1 z] - \partial_1 c + \beta \Delta \partial_1 \Theta= 0\ee 
Similarly the transformation of $B_2$ and $B_3$ yields

\begin{align}\begin{split} &2  [\partial_2 \chi \partial_0 a + \partial_0 z \partial_2 b - \partial_0 \chi \partial_2 a - \partial_2 z \partial_0 b] - \partial_2 c + \beta \Delta \partial_2 \Theta=0\\
&2 [ \partial_3 \chi \partial_0 a + \partial_0 z \partial_3 b - \partial_0 \chi \partial_3 b - \partial_3 z \partial_0 b] - \partial_3 c + \beta \Delta \partial_3 \Theta=0
\end{split}\end{align}
These can be written a single vector equation
\be\partial_0 f_i - \partial_i f_0- \partial_i c + \beta \Delta \partial_i \Theta = 0\ee

Using eq. (\ref{fv})  this can be written as:
\be \partial_i[\partial_0 \lambda - f_0 - c + \beta \Delta \Theta]=0 \label{c}\ee
yielding
\begin{align}\begin{split} c &= 2 (a \partial_0 \chi - b\partial_0 z) - \partial_0 \lambda - \beta \Delta \Theta \\
&=\beta\left[\frac{2}{E^2} \epsilon_{ijk}E_j  (\partial_0 \chi z_k - \partial_0 z \chi_k)\left[\Theta \hat n_i -\partial_i \int^x_\infty dl_C \hat{E}_l \hat n_l\Theta \right] +\partial_0 \int^x_\infty dl_C \hat{E}_i \hat n_i\Theta   - \Delta \Theta\right]
\label{c1}\end{split}\end{align}

Thus we find that the fields $z$, $\chi$ and $\Phi$ under Lorentz boost transform according to
eq.(\ref{LT})  with $a$,$b$,$c$ given in equations (\ref{ab}), (\ref{lambda}) and (\ref{c1}).

\section{Discussion}

In this note, we have amended the model  suggested previously as a candidate for effective  description of a gauge theory. We have considered only the abelian limit in which the model is equivalent to a theory of a free massless photon. We have proven this equivalence by considering the canonical structure of the theory. We have also shown that the basic fields of our model have interesting Lorentz transformation properties. 

We note that the modification discussed here also solves a certain puzzle posed by the suggestion of \cite{biz}. Namely, the model discussed in \cite{biz} possessed a global symmetry generated by
\begin{equation}\label{cf}
C_F=\int d^3x \left[p_z\frac{\partial G[z,\chi]}{\partial \chi}-p_\chi\frac{\partial G[z,\chi]}{\partial z}\right]
\end{equation} 
for an arbitrary function of two variables $G[z,\chi]$. These transformations constitute a group of area preserving diffeomorphisms on a sphere. The electric and magnetic fields $e$ and $b$ were invariant under the action of this group. That in fact suggested that the theory had some degrees of freedom in addition to the electrodynamic ones, since the action of the transformation generated by eq.(\ref{cf}) on the full phase space of the theory was nontrivial. However now the direct consequence of eqs. (\ref{pz}),(\ref{pchi}) and (\ref{fidot}), is that the generator of this transformation  vanishes,
\begin{equation}
C_F=0
\end{equation}
and thus there are no physical degrees of freedom that transform nontrivially under  eq.(\ref{cf}).
Thus in the present model the group of (global) diffeomorphisms $Sdiff(S^2)$ is in fact a global gauge symmetry, that is the states are invariant under the action of $C_F$, and what used to be extra degrees of freedom in \cite{biz} now becomes unphysical ``gauge'' coordinate. This ensures that electric and magnetic fields are the only physical degrees of freedom.

Since here we are dealing with the theory of a free photon, the charged states are not present. It should be however straightforward to extend this discussion to include electrically charged states. Just like in \cite{biz} we should lift the constraint of constant length of the field $\phi^a$, and instead allow dynamics of the modulus $\phi^2$. This will regulate the energy of the charged states in the UV and will make it finite. 
Since the configuration space of the model is $SO(3) \times R$, and the $SO(3)$ symmetry is broken to $O(2)$, the moduli space should have nontrivial homotopy group  $\Pi_2(M)= Z$, and the relevant topological charge should be identifiable with the electric charge\footnote{ There may be some subtlety in this argument related to the fact that the global gauge group $Sdiff(S^2)$ has to be modded out. However, since the gauge transformation is global, we do not anticipate any problems.}.

The next set of questions to be addressed is how to move in this theory to the nonabelian regime. According to the logic of \cite{biz} we need to find a perturbation that breaks the global symmetries of the model and through this breaking generates linear potential between the charges. The question one has to address, does this perturbation have to preserve the $Sdiff(S^2)$ gauge symmetry, or should it break it explicitly. This global gauge symmetry is a new element compared to 2+1 dimensions \cite{abelnonabel}, and we do not have any guidance from the 2+1 dimensional models.
Perhaps one should deal directly with the breaking of the generalized magnetic symmetry - the symmetry generated by the magnetic flux \cite{4d},\cite{generalized} in terms of its order parameter - the 't Hooft loop.\cite{zn}. These questions will be addressed in future work.


\begin{thebibliography}{1}
\bibitem{biz} I. B. Ilhan and A. Kovner; Phys.Rev. D88 (2013) 125004; e-Print: arXiv:1308.3865 [hep-th] 
	
\bibitem{zn} G. 't Hooft, Nucl.Phys. B138 (1978) 1  
	
\bibitem{abelnonabel} A. Kovner; e-Print: hep-ph/0009138, published in In Shifman, M. (ed.): At the frontier of particle physics, (World Scientific,
Singapore, 2001) vol. 3, 1777-1825; I. I. Kogan and A. Kovner; e-Print: hep-th/0205026, published in In *Shifman, M.
(ed.): At the frontier of particle physics, (World Scientific, Singapore, 2002) Vol. 4, 2335-2407

\bibitem{eduardo} E.I. Guendelman, E. Nissimov and S. Pacheva, Phys.Lett. B360 (1995) 57-64; e-Print: hep-th/9505128 

\bibitem{gliozzi} A somewhat similar model, but with a dual interpretation was considered in  Nucl. Phys. B141 (1978) 379.

\bibitem{4d} A. Kovner and B. Rosenstein, Phys.Rev. D49 (1994) 5571-5581; e-Print: hep-th/9210154 

\bibitem{generalized} D. Gaiotto, A. Kapustin, N. Seiberg and B. Willett, JHEP 1502 (2015) 172; e-Print: arXiv:1412.5148 [hep-th] 

\end{thebibliography}
\end{document}